\documentclass[twocolumn,twoside,9pt]{article}

\usepackage{ifpdf}
\ifpdf \usepackage[pdftex]{graphicx} \pdfcompresslevel=9
\else \usepackage[dvips]{graphicx} \fi

\usepackage{caption}
\usepackage{fancyhdr}
\usepackage{mathptmx}
\usepackage{graphicx}
\usepackage{times}

\usepackage[usenames]{color}
\usepackage[ruled,vlined]{algorithm2e}
\newcommand{\sfit}[1]{\textit{\sffamily #1}}

\usepackage[utf8x]{inputenc}
\usepackage{rotating} 
\usepackage{multirow} 

\title{Visualizing 2D Flows with Animated Arrow Plots}
\author{
Bruno Jobard$^{1}$, Nicolas Ray$^{2}$ and Dmitry Sokolov$^{3}$\\
~\\
$^1$~LIUPPA laboratory, University of Pau, France,
\textit{bruno.jobard@univ-pau.fr}\\
$^2$~ALICE Team, INRIA Nancy Grand-Est, France,
\textit{nicolas.ray@inria.fr}\\
$^3$~University of Lorraine, France,
\textit{dmitry.sokolov@loria.fr}
}
\date{}

\begin{document}

\twocolumn[{
\renewcommand\twocolumn[1][]{#1}

\maketitle

\begin{center}
\centering
\includegraphics[width=\linewidth]{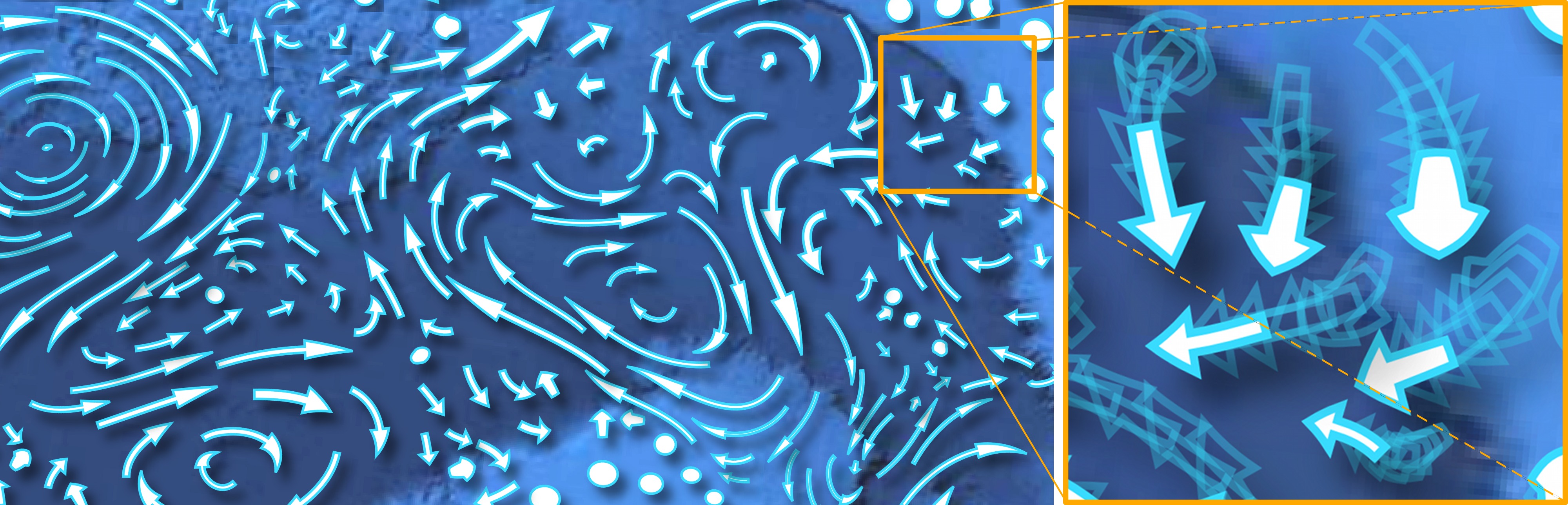}
\captionof{figure}{\textbf{Ocean currents visualized with a set of dynamic arrows.}
(Left) The domain is filled with arrows aligned with the flow. The length is
proportional to the velocity magnitude. The arrow density is controlled by a
custom map to better capture local turbulences. (Right) Close-up showing the
arrow trajectories and the morphing of their glyphs.}
\label{fig:teaser}
\end{center}
}] 

\begin{abstract}
Flow fields are often represented by a set of static arrows to illustrate
scientific vulgarization, documentary film, meteorology, etc. This simple
schematic represention lets an observer intuitively interpret the main
properties of a flow: its orientation and velocity magnitude. We
propose to generate dynamic versions of such representations for $2D$
unsteady flow fields. Our algorithm smoothly animates arrows along the flow
while controlling their density in the domain over time. Several
strategies have been combined to lower the unavoidable popping artifacts arising
when arrows appear and disappear and to achieve visually pleasing animations.
Disturbing arrow rotations in low velocity regions are also handled by
continuously morphing arrow glyphs to semi-transparent discs.
To substantiate our method, we provide results for synthetic and real velocity
field datasets.

\end{abstract}

\section*{Introduction}

Arrow plots are standard static representations for 2D vector fields. They are
intuitive and thus often used to present flows mixed with a contextual
background image to non-expert public. The goal of this work is to provide a
simple algorithm that produces clean animated arrow plots for presentation
purpose.

This aim greatly differs from the traditional objective as evidenced by the most
recent 2D vector field visualization techniques, where the efforts have focused
on the interactive exploration of the data. For the purpose of exploration,
image-based techniques such as flow textures (LIC and its animated extensions)
allow for interactive visualization of flow details by using every pixels of the
display device to communicate dense information. However, flow textures present
two major drawbacks in our context: blending them with an additional color map
(a background image or a dynamic field) might greatly deteriorate its details,
and such representations are quite sensitive to the quality of the display
device and cannot be controlled while broadcasting videos or during a public
presentation. For the purpose of presentation, a sparse set of moving arrows can
easily convey the desired information. The arrows do not deteriorate the
background information (only occlude it temporary), and is robust to low quality
display device.

This work proposes a method producing sparse and smoothly animated
representations of a flow with moving arrows (Figure~\ref{fig:teaser}). We list
below the general properties and constraints such an algorithm should satisfy.

\textbf{Arrow trajectories:} to intuitively convey its dynamic nature,
the arrow trajectories should follow the flow.

\textbf{Local flow depiction:} the arrow shape should depict the local
orientation and velocity magnitude of the flow at any time.

\textbf{Uniform domain coverage:} the representation should provide an
uncluttered information of the flow everywhere in the domain at any time.

\textbf{Smooth animations:} the arrow movement should be as smooth as
possible to avoid distraction.

We propose an algorithm that generates intuitive arrow plot animations by
advecting and bending arrows over time while guaranteeing that arrows will not
occlude each other and ensuring a complete coverage of the domain. Moreover, the
method is able to adapt the density of arrows to arbitrary density field.

However, keeping a uniform coverage of the domain with no occlusion involves
inserting new arrows to fill the empty places and removing some arrows in places
that get too crowded. This necessary insertion and deletion of arrows introduce
strong popping artifacts when they appear and disappear, deteriorates the
smoothness of the animation. Our algorithm has been designed to minimize it,
both when generating arrows and at rendering time.

The main contributions in this paper are :
\begin{itemize}
  \item an efficient algorithm that controls the density of arrows and manages
  their life span while maintaining low popping artifacts,
  \item a rendering algorithm that further reduces popping by both fading
  arrows in and out, and a morphing strategy that handles transitions
  between high and low velocity regions,
  \item and experimentations on both real and synthetic flows to evaluate how
  much the (unavoidable) popping artifacts can be maintained acceptable for
  visualization purposes.
\end{itemize}

The rest of the paper is organized as follows: after reviewing the state of the
art, our arrow representation is introduced (Section~\ref{sec:pbsettings}), the
arrow generation algorithm is provided (Section~\ref{sec:generate-arrows}),
the capacity to adapt the arrow density to any
density field is explained (Section~\ref{sec:variable-density}), the rendering
method is presented (Section~\ref{sec:rendering_arrows}), and results are presented
(Section~\ref{sec:results}) and discussed (Section~\ref{sec:discussion}).

\section*{Related work}

Most of the recent work on visualizing 2D vector field have targeted the ability
for scientists to interactively explore their flow datasets. This task is very
efficiently achieved with texture-based techniques~\cite{laramee_state_2004} which
offer a dense representation of the fine details of a vector field. They are
inexpensive to compute and can produce smooth animations of unsteady flows.
However, when combined with a background image, these texture-based
representations sometimes fail in displaying both the fine details of the flow
and the background (see Section~\ref{sec:comparison_previous_works} and
Figure~\ref{fig:comparison_lic}). In the context of presenting flow behavior in
an animated way to non-expert public, simpler and more schematic alternatives
are more attractive.

Numerous such geometric based flow visualization methods were invented over the
two last decades~\cite{mcloughlin_over_2010}. We focus below on the techniques
closely related to visualization of a flow field by animated geometric
primitives.

\textbf{Vector plots:} The simplest vector field visualization method consists
in drawing straight segments originating from the nodes of an underlying mesh
\cite{dovey_vector_1995} (Possibly a Cartesien grid) to indicate the local flow
direction and possibly its orientation by placing an arrow tip at its other end.
Its magnitude might be conveyed by the segment length. The main drawback comes
from the origin of arrows being unable to change over time, leading to
occlusions~\cite{klassen_shadowed_1991} and confusing animations when the vector
field is a flow.

\textbf{Arrow placement:} In flow visualization, clutter and occlusion problems
have been mainly addressed in the context of streamline placement methods. These
algorithms apply here since an arrow can be carried by a small streamline
(streamlet) to better depict the local flow. Any of these numerous methods
\cite{turk_image-guided_1996,jobard_creating_1997,mebarki_farthest_2005,liu_advanced_2006}
can be used since they guarantee that no streamlet will be placed within a
distance $d_{sep}$ to its neighbours. It is also possible to adapt the
streamline density~\cite{schlemmer_priority_2007}. An animated streamline
placement has been proposed by Jobard and Lefer \cite{jobard_unsteady_2000}.
This later work renders the streamlines with an animated texture that looks like
particle trails advected along the streamline. Restricting these particle trails
to be aligned on streamlines makes it impossible to avdect them in the flow, and
constrains all particles of a streamline to be born and die together. Moreover,
the lifetime of streamlines is more sensitive to flow evolution e.g. a flow with
constant rotation in time will create spinning arrows with our algorithm whereas
streamlines would have very short lifetime.

Other methods have been proposed to nicely distribute glyphs. Hiller {\it et
al.} \cite{hiller_beyond_2003} minimizes Lloyd's energy to
evenly distribute glyph's positions and other works aim to place a minimal
number of glyphs~\cite{telea_simplified_1999, mckenzie_vector_2005} to represent
the flow. However, these works do not extend nicely to unsteady flows. An error
diffusion approach has been proposed \cite{hausner_animated_2006} to distribute
glyphs in unsteady flows, but it exhibits both high popping and the distribution
is not convincing.

\textbf{Particle tracing:} The dynamics of the flow can be revealed by
visualizing particles advected in the domain. Contrary to arrows, the small size
of particle glyphs minimizes the occlusion problems. Inter particle distances
has not to be checked and the density is mainly controlled by the seeding
strategy. Bauer et al. use tiles of Sobol quasi-random positions to regularly
seed particles in region of interest of unsteady 3D flows
\cite{bauer_case_2002}. Since they deal with incompressible flows, the initially
constant density of injected particles remains constant during the advection
process. Following the same framework, Helgeland and
Elboth~\cite{helgeland_high-quality_2006} enhanced the rendering with
anisotropic diffusion to better represent the flow. It is even possible to
represent particules with arrows that are advected by the
flow~\cite{van_wijk_image_2002}. However, arrows will suffer shearing and could
only be used when the support is given by a pathlet i.e. advecting a streamlet do
not produce a streamlet at the new frame.

Our algorithm better covers the domain and avoids arrow overlaps thanks to more complex
creation and deletion strategies, including backward propagation. Somehow, it requires
relaxing the realtime feature of particle tracing.

\section{Moving arrow representation}
\label{sec:pbsettings}

\begin{figure}[!htb]
\centering
\includegraphics[width=0.7\linewidth]{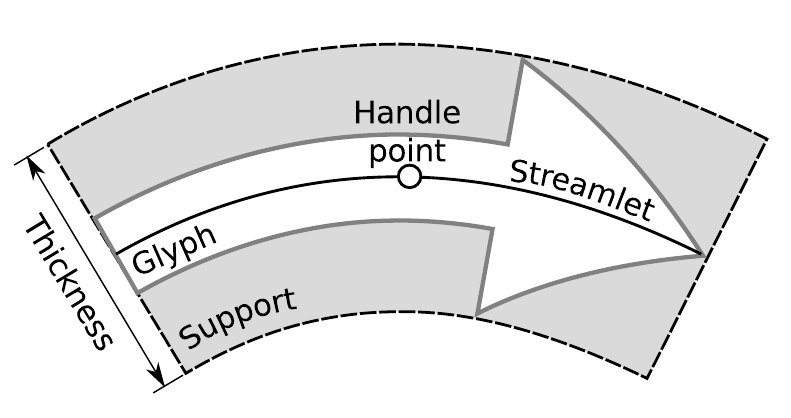}
\caption{\textbf{Arrow Anatomy.} An arrow glyph is mapped on a rectangular
support of a given thickness warped along a streamlet integrated from a central handle
point.}
\label{fig:anatomy}
\end{figure}

During the animation, the arrows are represented with glyphs mapped on
rectangular supports. The supports are warped according to the local flow
orientation and magitude. Each arrow is initiated from a so-called
\textit{handle point} and its support is then warped according to a short
streamline integrated backward and forward from the handle point (see
figure~\ref{fig:anatomy}). The integration length of the streamlets is such that
their length is proportional to the local velocity magnitude of the flow.

More formally, given a 2D time-dependent vector field
$\mathbf{v}(\mathbf{x},t)=(v_x, v_y)$, a streamline $S$ is a parametric curve
$S(\tau)$ defined at time $t$ and initiated from an handle point $p$. $S(\tau)$
is given by the equation: $$\frac{dS}{d\tau}=\mathbf{v}(S(\tau),t)
\mathrm{~with~} S(0)=p$$

The streamlets have a constant integration length $L$ and any sample point of
the streamlet $S$ is given by: $$S(l) = S(0) +
\int_{0}^{l}\mathbf{v}(S(\tau),t)d\tau \mathrm{~with~} l\in [-\frac{L}{2},
\frac{L}{2}]$$ A standard Runge-Kutta integration scheme is used to sample the
streamlets backward and forward from their handle point.

To prevent having severely distorted arrows in high velocity regions, streamlets
might be clamped if the support aspect ratio (arrow length over thickness)
becomes superior to a given user threshold.

To intuitively convey the dynamics of the flow, it is preferable for the arrows
to be transported along the flow. Since their shape is determined at any time step by
the streamlet integration, only the handle point is advected from one step to
the next. This particular point follows a pathline trajectory $P$ initiated
from a seed point $p_s$ at time $t_s$: $$\frac{dP}{dt}=\mathbf{v}(P(t),t)
\mathrm{~with~} P(t_s)=p_s$$

An arrow will then move along this trajectory between its birth time $t_b$ and
its death time $t_d$ ($t_b \leq t_s \leq t_d$).  Both $t_b$ and $t_d$ are determined
by the arrow reaching the boundaries of the space-time domain or by a lack of empty
space requiered to place its support as discussed in the following section.

\section{Uniform placement of moving arrows}
\label{sec:generate-arrows}

To uniformly place moving arrows over the domain, our algorithm successively
fills each time step with as many arrows as possible such that a minimal
"separating" distance $d_{sep}$ is respected between arrows. The $d_{sep}$
parameter controls the tradeoff between the competing objectives of
avoiding cluttering and covering the whole domain.

Animation frames are populated with arrows by the
Algorithm~\ref{alg:build_arrow_trajectories}, which consists in three stages.

\begin{figure}[!htb]
\centering
\includegraphics[width=.75\linewidth]{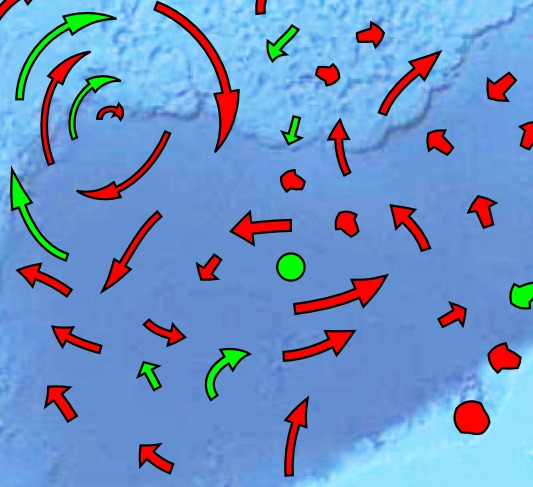}\\
\vspace{1mm}
\includegraphics[width=.75\linewidth]{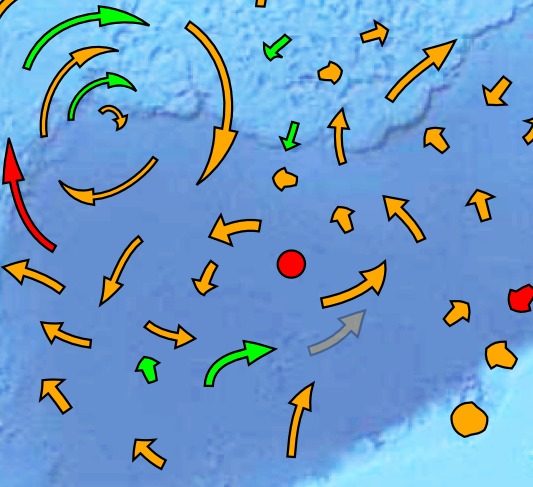}
\caption{\textbf{First and second time step of an animation.} (Top) The empty
domain is first filled with $d_{seed}$-separated arrows (red). (Bottom) The
arrows from the previous time step are propagated into the current one (orange)
and new arrows are inserted. (Right to left) Then empty spaces are filled with
backward propagated arrows (green) from future time steps. The semi-transparent
arrow on the right image is removed to preserve the minimal separating
distance.}
\label{fig:CompleteTimeStepWithArrows}
\end{figure}

\begin{itemize}
\item \textbf{\textsf{S1}.} The first stage fills the first time step with
evenly-spaced arrows (see Section \ref{sec:unif-arrow-placement} and the red
arrows in Figure~\ref{fig:CompleteTimeStepWithArrows}).

\item \textbf{\textsf{S2}.} The second stage iterates over the time steps.
First, the arrows that can be propagated from the previous time step are
inserted into the current one (see Section \ref{sec:advect-one-step} and the
orange arrows in Figure~\ref{fig:CompleteTimeStepWithArrows}). Second, the
current time step is completed with new arrows inserted in the previous stage
(see the red arrows in Figure~\ref{fig:CompleteTimeStepWithArrows}, bottom).

\item \textbf{\textsf{S3}.} The third stage iterates from the last time step to
the first one, and advances arrow's birth when possible, thus increasing arrow's
lifetime (see the ``backward'' paragraph in Section \ref{sec:advect-one-step}
and the green arrows in Figure~\ref{fig:CompleteTimeStepWithArrows}).
\end{itemize}

\begin{algorithm}
\SetFuncSty{sfit}
\SetKwSty{sfbf}
\SetArgSty{}
\SetCommentSty{itshape}
\SetKwInOut{KwIn}{Input}
\SetKwInOut{KwOut}{Output}
\SetKwInOut{KwInOut}{In/Out}
\SetKwInOut{KwData}{Data}
\SetKwData{vf}{vf}
\SetKwData{tmax}{tmax}
\SetKwData{gTraj}{arrows}
\SetKwData{Time}{time}
\SetKwData{forward}{``forward''}
\SetKwData{backward}{``backward''}
\SetKwData{direction}{$\Delta t$}
\SetKwFunction{CompleteTimeStepWithArrows}{CompleteTimeStepWithArrows}
\SetKwFunction{AdvectArrows}{PropagateArrowsOneStep}
\SetKw{KwDownTo}{downTo}

\hspace{-4.5mm}\begin{tabular}{ll}
\KwOut{\gTraj} & \tcp{set of arrows} \\
\KwData{\tmax} & \tcp{max vector field time step}
\end{tabular}
\BlankLine

\hspace{-4.5mm}\begin{tabular}{ll}
\multirow{1}{*}{\begin{sideways}\textbf{\textsf{S1}}\end{sideways}} &
\begin{minipage}[b]{0.85\linewidth}
\CompleteTimeStepWithArrows($0$, \gTraj)\;
\end{minipage}\\
\multirow{1}{*}{\begin{sideways}\parbox{7mm}{\textbf{\textsf{S2}}}\end{sideways}}
&
\begin{minipage}[b]{0.85\linewidth}
\For{\Time $\leftarrow 1$ \KwTo \tmax}{
	\direction $\leftarrow 1$; \tcp*[f]{forward advection}\\
	\AdvectArrows(\Time, \direction, \gTraj)\;
	\CompleteTimeStepWithArrows(\Time, \gTraj)\;
	}
\end{minipage}\\
\multirow{1}{*}{\begin{sideways}\parbox{5mm}{\textbf{\textsf{S3}}}\end{sideways}}
&
\begin{minipage}[b]{0.85\linewidth}
\For{\Time $\leftarrow$ \tmax $-~1$  \KwDownTo $0$}{
	\direction $\leftarrow -1$; \tcp*[f]{backward advection}\\
	\AdvectArrows(\Time, \direction, \gTraj)\;
	}
\end{minipage}
\end{tabular}

\caption{\sfit{PlaceMovingArrows}}
\label{alg:build_arrow_trajectories}
\end{algorithm}

The two next sections explain how individual animation frames are populated with
new arrows (see Section~\ref{sec:unif-arrow-placement}) and how these arrows
will be propagated to first populate the next time step (see
Section~\ref{sec:advect-one-step}).

\subsection{Completing a time step with evenly-spaced arrows}
\label{sec:unif-arrow-placement}

Evenly distributing arrows over a $2D$ domain could be addressed by Lloyd's
relaxation~\cite{hiller_beyond_2003}. However, in the dynamic case, it is
sufficient to use a faster algorithm as the distribution quality will decrease
rapidly due to arrow displacements. Therefore the problem can be reduced here to
the well studied placement of streamlets. We implemented a quite standard greedy
approach that works as follows (see Algorithm~\ref{alg:fill_slice}): from a
sufficiently dense sampling of the domain, streamlets are successively
integrated from these sample positions and are inserted into the representation
if their distance to already placed streamlets is superior to a seeding distance
$d_{seed}$ (with $d_{seed}>d_{sep}$ as explained in
Section~\ref{sec:next-arrow}).

\begin{algorithm}
\SetFuncSty{sfit}
\SetKwSty{sfbf}
\SetArgSty{}
\SetCommentSty{itshape}
\SetKwInOut{KwIn}{Input}
\SetKwInOut{KwOut}{Output}
\SetKwInOut{KwInOut}{In/Out}
\SetKwInOut{KwData}{Data}
\SetKwData{vf}{vf}
\SetKwData{tmax}{tmax}
\SetKwData{gTraj}{arrows}
\SetKwData{dseed}{$\mathsf{d_{seed}}$}
\SetKwData{Time}{time}
\SetKwData{seedPos}{seedPositions}
\SetKwData{pos}{position}
\SetKwData{support}{support}
\SetKwData{arrow}{newArrow}
\SetKwData{traj}{trajectory}
\SetKwFunction{CreateTraj}{CreateTraj}
\SetKwFunction{CreateSupport}{CreateSupport}
\SetKwFunction{CreateArrow}{CreateArrow}
\SetKwFunction{Distance}{Distance}
\SetKwFunction{CreateTraj}{CreateTraj}
\SetKwFunction{Insert}{Insert}

\hspace{-4.5mm}\begin{tabular}{ll}
\KwIn{\Time}     & \tcp{current time step} \\
\KwInOut{\gTraj} & \tcp{set of arrows} \\
\KwData{\dseed}  & \tcp{seeding distance}
\end{tabular}
\BlankLine
Fill a vector \seedPos with domain sampling positions\;
\ForEach{\pos $\in$ \seedPos}{
\arrow $\leftarrow$ \CreateArrow(\pos, \Time)\;
\If{\Distance(\arrow, \gTraj, \Time)$>$ \dseed}{
	\gTraj.\Insert(\arrow)\;
	}
}
\caption{\sfit{CompleteTimeStepWithArrows}}
\label{alg:fill_slice}
\end{algorithm}

Algorithm~\ref{alg:fill_slice} requires evaluation of distance
between a new candidate arrow and the previously placed arrows
(Section~\ref{sec:arrow_dist}), and a seeding strategy that determines where to
place candidate arrows (Section~\ref{sec:next-arrow}) and when to stop trying.

\subsubsection{Evaluating the distance between arrows}
\label{sec:arrow_dist}

The distance between arrows is approximated by the minimal distance between
their streamlets. In our implementation, the distance from any point of the
domain to the existing arrows is stored in a discretized distance map, which is
filled with a fast marching algorithm. The distance requests are then fast to
process since they only require accessing the distance map at the requested
locations. The distance map resolution is defined with respect to $d_{sep}$ as
illustrated in Figure~\ref{fig:resolution}. This approach also remains efficient
in the presence of adaptive arrow density (see
Section~\ref{sec-raster-dist-field}).

\subsubsection{Choosing where to seed the arrows}
\label{sec:next-arrow}

The purpose of the seeding strategy is to cover the whole domain with arrows as
close as possible to the maximum authorized density. A simple solution is to
create a shuffled list of positions in a Cartesian grid, and always select the
next position in the list.

In Algorithm~\ref{alg:fill_slice} the arrow density is controlled by the
parameter $d_{seed}$, which would be equal to $d_{sep}$ for static
representations. Since arrows will be propagated to the next step (see
Section~\ref{sec:advect-one-step}), we introduce a security distance by taking
$d_{seed}>d_{sep}$. This way, individual arrows have a higher probability to be
propagated more steps ahead before their distance to surrounding arrows falls
under the $d_{sep}$ threshold. We observe that setting $d_{seed}= 2 d_{sep}$ leads
to satisfying results. The ratio $d_{seed}/d_{sep}$ manages the tradeoff between
the arrows life-time and the uniform domain coverage.

\subsection{Arrow propagation to the next step}
\label{sec:advect-one-step}

For each time step $t$, new arrows are iteratively introduced by
Algorithm~\ref{alg:fill_slice} but first, it is tested if it is possible to
increase the life span of the arrows that exist at time step $t-1$ (forward) or $t+1$
(backward) with Algorithm~\ref{alg:advect_one_step}.

\begin{figure}[!htb]
\centering
\includegraphics[width=.455\linewidth]{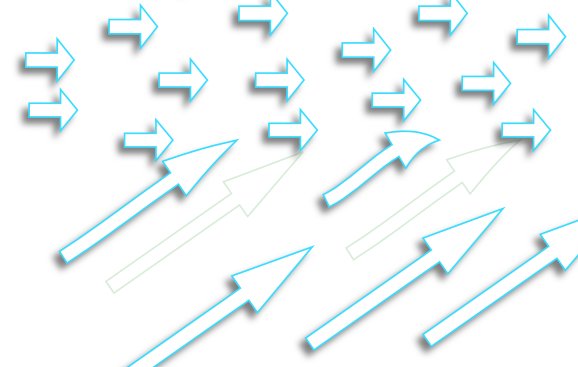}
\hspace{0.5cm}
\includegraphics[width=.455\linewidth]{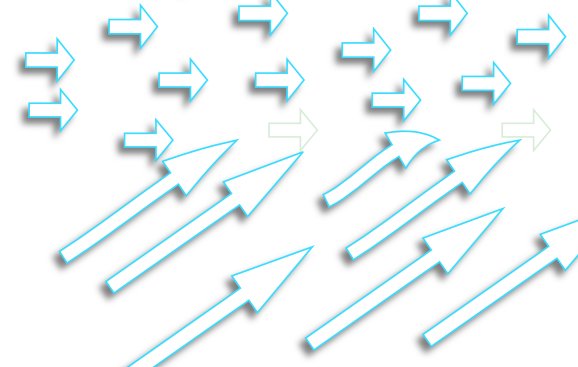}
\caption{\textbf{Arrow propagation with priority.} Giving priority to short
arrows (left) would kill long arrows and therefore waste a lot of space and
create noticeable popping artifacts. Our strategy to propagate long arrows first (right)
resolves this issue by removing small arrows.}
\label{fig:priority}
\end{figure}

\textbf{During the forward stage}, arrows evolve through advection (of its handle
point) and may enter into conflict with one another. Therefore, some arrows are
to be deleted and the choice is made by a greedy approach: all arrows alive at
time $t-1$ are taken in a heuristic order and propagated one by one to time $t$.
If an arrows conflicts with already propagated arrows, it is discarded. This
method is fast and allows to favor arrows by defining priority criteria. In
practice, sorting arrows by decreasing streamlet length in screen space allows
to keep long arrows as much as possible, and therefore minimizes the popping
artifacts. The benefits of this strategy are illustrated in
Figure~\ref{fig:priority}.

\textbf{The backward stage} is similar to the forward one, except that it does not
try to introduce new arrows because the empty spaces have already been filled
during the forward stage. However, since the arrows have been seeded at least at
the $d_{seed}$ distance to the surrounding ones, some of these arrows might find
enough place to propagate back until they reach the $d_{sep}$ separating
distance. These cases are illustrated with the green arrows on
Figure~\ref{fig:CompleteTimeStepWithArrows}.

To obtain progessive appearance and disappearance of the arrows along the
borders of the frame, it is necessary to extend the domain with a buffer zone
where the vector field is extrapolated. The size of this hidden buffer zone is
related to half the maximal length of the arrows. All the operations are
performed on this extended domain.

\begin{algorithm}
\SetFuncSty{sfit}
\SetKwSty{sfbf}
\SetArgSty{}
\SetCommentSty{itshape}
\SetProcNameSty{sfit}
\SetKwInOut{KwIn}{Input}
\SetKwInOut{KwOut}{Output}
\SetKwInOut{KwInOut}{In/Out}
\SetKwInOut{KwData}{Data}
\SetKwData{vf}{vf}
\SetKwData{gTraj}{arrows}
\SetKwData{direction}{$\Delta t$}
\SetKwData{forward}{``forward''}
\SetKwData{backward}{``backward''}
\SetKwData{Time}{time}
\SetKwData{prevTime}{prevTime}
\SetKwData{support}{support}
\SetKwData{arrow}{arrow}
\SetKwData{dsep}{$\mathsf{d_{sep}}$}
\SetKwData{traj}{trajectory}
\SetKwData{trajToGrow}{growingCandidates}
\SetKwData{curPos}{curPos}
\SetKwData{newPos}{newPos}
\SetKwFunction{CreateSupport}{CreateSupport}
\SetKwFunction{CreateArrow}{CreateArrow}
\SetKwFunction{Distance}{Distance}
\SetKwFunction{Insert}{Insert}
\SetKwFunction{CanGrow}{IsAlive}
\SetKwFunction{EndPoint}{EndPoint}
\SetKwFunction{Integrate}{Integrate}
\SetKwFunction{CannotGrow}{CannotGrow}
\SetKwFunction{Sort}{SortByPriority}
\SetKwFunction{Arrows}{Arrows}
\SetKwFunction{propagateto}{PropagateTo}
\SetKwFunction{RemoveTimeStep}{RemoveTimeStep}

\hspace{-4.5mm}\begin{tabular}{ll}
\KwIn{\Time}      & \tcp{current time step} \\
\KwIn{\direction} & \tcp{1 is forward, -1 is backward} \\
\KwInOut{\gTraj}  & \tcp{set of arrows} \\
\KwData{\vf}      & \tcp{velocity vector field} \\
\KwData{\dsep}    & \tcp{separating distance} \\
\end{tabular}
\BlankLine
\prevTime $\leftarrow$ \Time $-$ \direction\;
\trajToGrow $\leftarrow \emptyset$; \tcp*[f]{empty list}\\
\ForEach{\arrow $\in$ \gTraj}{
	\If{\arrow.\CanGrow(\prevTime) \\
	\textbf{and not} \arrow.\CanGrow (\Time) }{
		\trajToGrow.\Insert(\arrow)\;
	}
}
\Sort(\trajToGrow)\;
\ForEach{\arrow $\in$ \trajToGrow}{
\arrow .\propagateto (\Time)\;
	\If{\Distance(\arrow, \gTraj, \Time)$<$ \dsep}{
	\arrow .\RemoveTimeStep(\Time)\;
	}
}
\caption{\sfit{PropagateArrowsOneStep}}
\label{alg:advect_one_step}
\end{algorithm}

\section{Introducing an adaptive density of arrows}
\label{sec:variable-density}

\begin{figure}[!h]
\centering
\includegraphics[angle=-90,width=.49\linewidth]{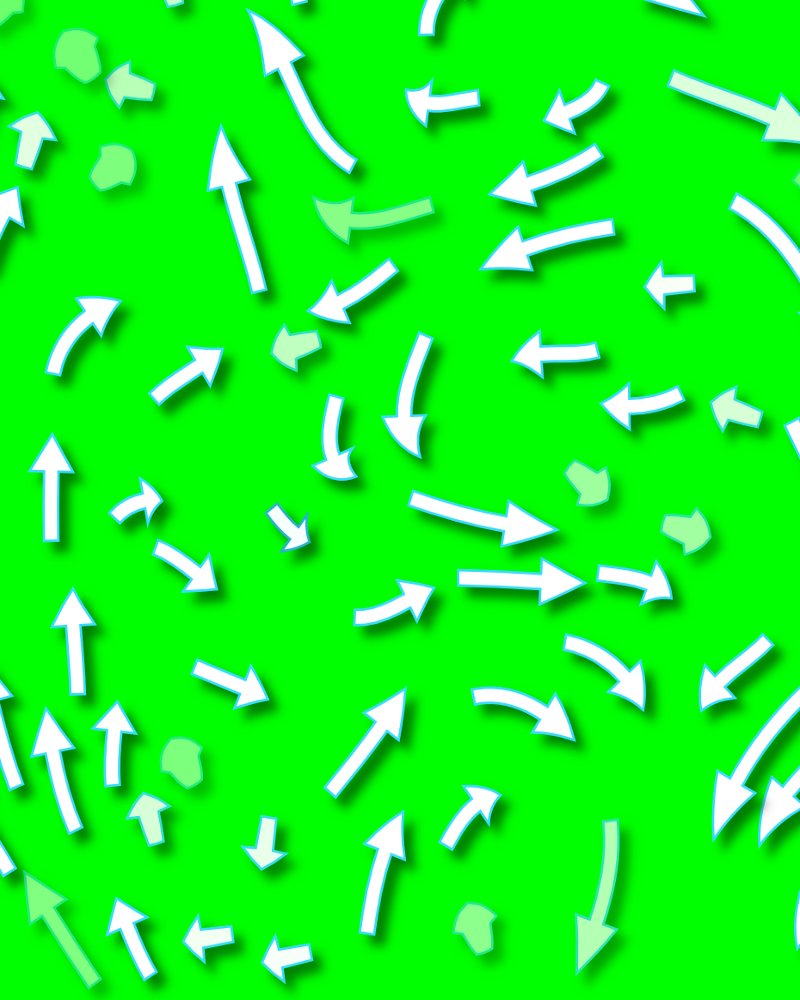}
\includegraphics[angle=-90,width=.49\linewidth]{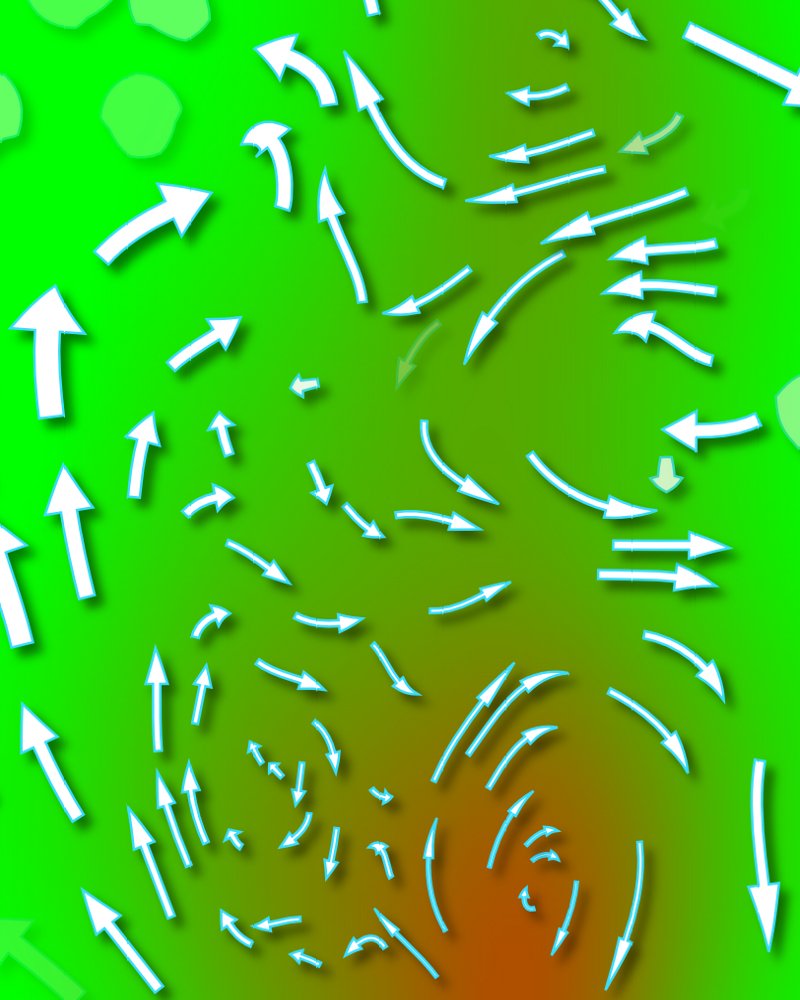}
\caption{\textbf{Uniform vs. adaptive density of arrows.} (Left) A uniform
density of arrows will fail at revealing the turbulent areas if the separating
distance is too high -- or would overpopulate the domain with a small separating
distance. (Right) Adapting the separating distance to a density field (green to
red background) will adapt the number of arrows necessary for depicting the details
of the flow.}
\label{fig:density}
\end{figure}

The algorithm defined in the previous section computes a sparse set of moving
arrows that keeps a uniform density over time. However, it is often interesting
to adapt the density of arrows to the local features of the flow (see
Figure~\ref{fig:density}). This can be done by introducing a density map which
measures the scale at which the phenomenon needs to be captured.

The density map is a scalar field that gives the local "zoom factor" of the
field: in particular, a value of $1$ means that the arrows are generated as with
the previous algorithm, and in general a value $scale$ means that the algorithm
generates arrows as such that a close-up (of factor $scale$) of the region gives
the same appearance as the previous algorithm. We first discuss possible ways
to automatically define the density map, then present the modifications of the
algorithm required to add this feature.

\subsection{Choice of a density map}
\label{sec:dens-fct}

The density map can be used to add more details where the flow has more
variations. It is therefore natural to estimate it by a differential quantity
derived from the vector field. In our experiments, we estimate it by the
Frobenius norm of the Jacobian matrix i.e. $\sqrt{({\partial v_x/\partial
x})^2+({\partial v_x/\partial y})^2 +({\partial v_y/\partial x})^2+({\partial
v_y/\partial y})^2}$, at the position $x,y$ with the velocity field
$v(v_x,v_y)$. This quantity allows to focus on flow features as it is correlated with
both divergence and curl. However, other density maps may be more appropriate in
particular cases, such as the velocity or vorticity magnitude as discussed in
Schlemmer \textit{et al.}'s work \cite{schlemmer_priority_2007}. Regardless the way
for estimating the density map, its values are set in the range
$[1,scale_{max}]$ so that some regions can exhibit an arrow placement
$scale_{max}$ times denser than others. Most frequently we set $scale_{max}\leq
10$.

\subsection{Adapting the algorithm to handle a density map}
\label{sec-raster-dist-field}

The only thing we need to change in Algorithm~\ref{alg:build_arrow_trajectories}
in order to take the density map into account is the distance
between arrows: the new distance is just scaled by the density. When comparing
this new distance with $d_{sep}$ and $d_{seed}$ in the algorithm, the spacing
between neighboring arrows becomes proportional to the inverse of the density.

The new distance is therefore the weigthed distance with respect to the density.
A weigthed distance is formaly defined in \cite{memoli_fast_2001}, and is
commonly used in images applications such as image segmentation
\cite{protiere_interactive_2007}.

As introduced in section \ref{sec:arrow_dist} the distance of each point to
previously placed arrows is stored in a distance map. The distance map is
now considered as a weighted graph where each pixel is connected to its $8$
closest neighbors and the weights corresponds to the edge length ($1$ or
$\sqrt{2}$ for diagonals) times the density map evaluated at this position. The
distance between two points (pixels) is then the cost of the shortest path in
this graph.

Evaluating the distance to a new arrow from all previously placed arrows only
requires reading the distance map at the sampling points of the new arrow.

Updating the distance map is a bit more difficult. It first requires to be
initialized to $\infty$, then for each new inserted arrow, the distance field is
updated by a $n$-seed Dijkstra algorithm (placing one seed for each point
touched by a rasterization of the arrow's streamlet). The very specific nature
of the graph (bounded weights between $1$ and $\sqrt{2} \times scale_{max}$)
makes it possible to use simpler and more efficient algorithms such as
\cite{yatziv_o_2006}.

\begin{figure}[!t]
\centering
\includegraphics[width=.8\linewidth]{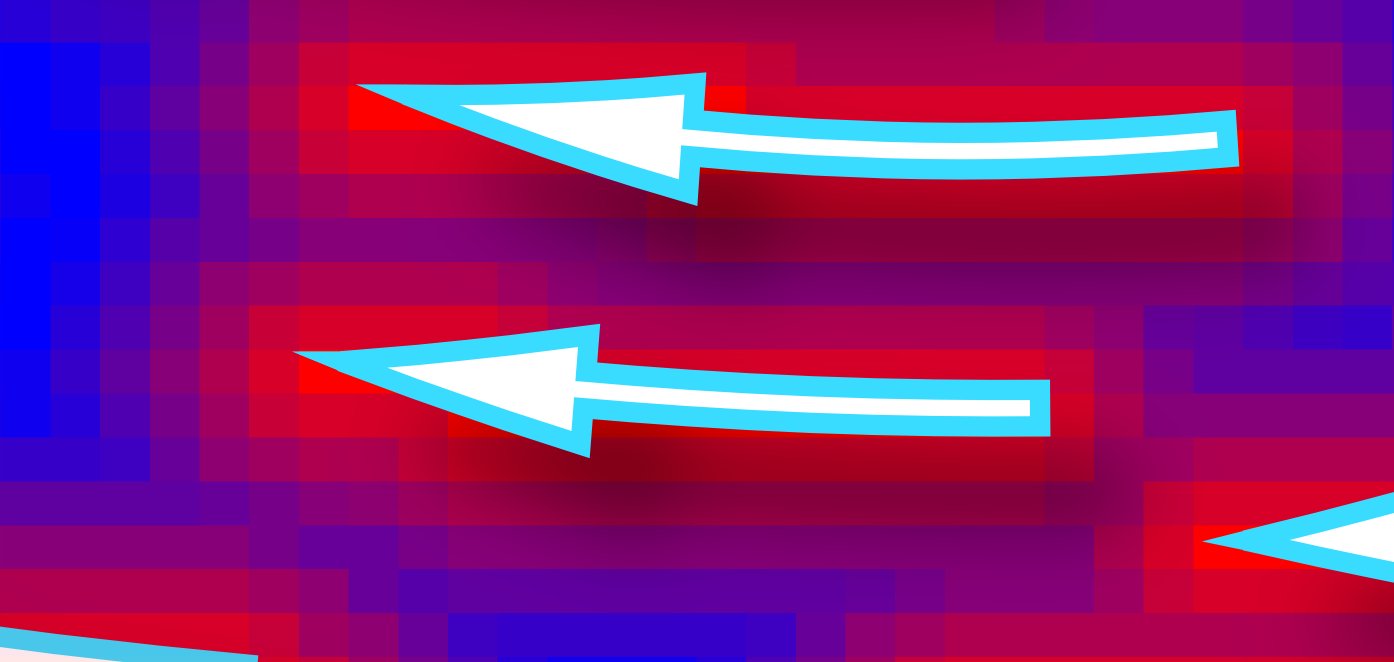}
\caption{\textbf{Resolution of the distance map.} Setting the pixel width to be
a quarter of $d_{sep}$ achieves a good trade-off between the update time of the distance
map and its accuracy.}
\label{fig:resolution}
\end{figure}

As illustrated in Figure~\ref{fig:resolution}, setting the pixel size of the
distance map to be a quarter of the separating distance is enough to discretize
the distance map.

\section{Rendering arrows}
\label{sec:rendering_arrows}

To draw all the arrows, the rendering algorithm determines the mapping of the
arrow glyph onto the screen (Section~\ref{sec:mapping}), morphs the arrows to
new symbols when the flow magnitude becomes too low
(Section~\ref{sec:morphing}), and adds transparency to reduce the popping
artifacts (Section~\ref{sec:popping}).

\subsection{Arrow mapping}
\label{sec:mapping}

The glyphs are centered on the handle point of the arrows. Our algorithm
precomputes these positions at each time step. For frames in-between time steps,
the smoothness of arrow displacements is ensured by a cubic hermite
interpolation (in time) of the handle point. From the interpolated handle
points, a streamlet is integrated and a support is computed by thickenning it.
For adaptive density, the thickness of the support is divided by the local
density to avoid occlusion with closest arrows.

Notice that it would be possible to integrate (in time) the handle point
position with Runge--Kutta, but we prefer to interpolate them due to the
asymmetry of the integration scheme.

\subsection{Fade-in and fade-out of arrows}
\label{sec:popping}

Popping effects come generally from the insertion and deletion of arrows. To
attenuate this effect, the arrows are rendered with an opacity coefficient that
smoothly decreases near birth and death time steps. When the number of rendering
frames is higher than the simulation time step, all arrows born (resp. died) at
time $t$ have the same transparency coefficient, leading to a visual artifact.
This can be solved by adding a random delay (shorter than a time step) before
applying the fading effect.

\subsection{Arrow morphing}
\label{sec:morphing}

Arrows are standard symbols to represent flows, however it may become misleading
where the flow magnitude becomes too low. Inspired by meteorologists, we decided
to draw discs to indicate calm regions.

\begin{figure}[!htb]
\centering
\includegraphics[width=.99\linewidth]{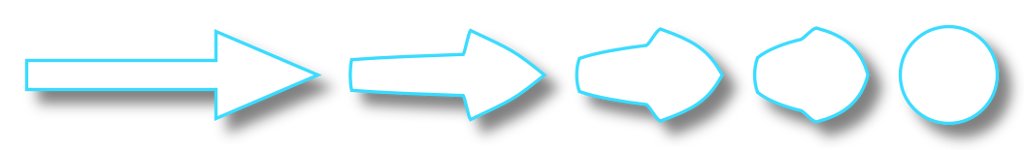}
\caption{\textbf{Glyph morphing sequence.} The arrows are smoothly morphed to
discs in regions of low velocity.}
\label{fig:morph}
\end{figure}

To render dynamic flows, a smooth transition between arrows and discs (see
Figure~\ref{fig:morph}) allows the avoidance of
distracting symbol switches. The glyph to use is determined by the streamlet
length over glyph thickness ratio: this ensures that arrows length is always
greater than arrows thickness.

The support of the glyph also requires a special treatment when its streamlet is
shorter than the arrows thickness: the support is no longer obtained by
thickening the streamlet, but by drawing a square centered on the handle point
and oriented by the vector between the streamlet extremities.

Using a symbol with rotational invariance (disc) prevents the user from being
distracted by the frequent rotations when the flow velocity is almost null.

\section{Results}
\label{sec:results}

\subsection*{Real datas}

\begin{figure*}[!t]
\centering
\includegraphics[width=\linewidth]{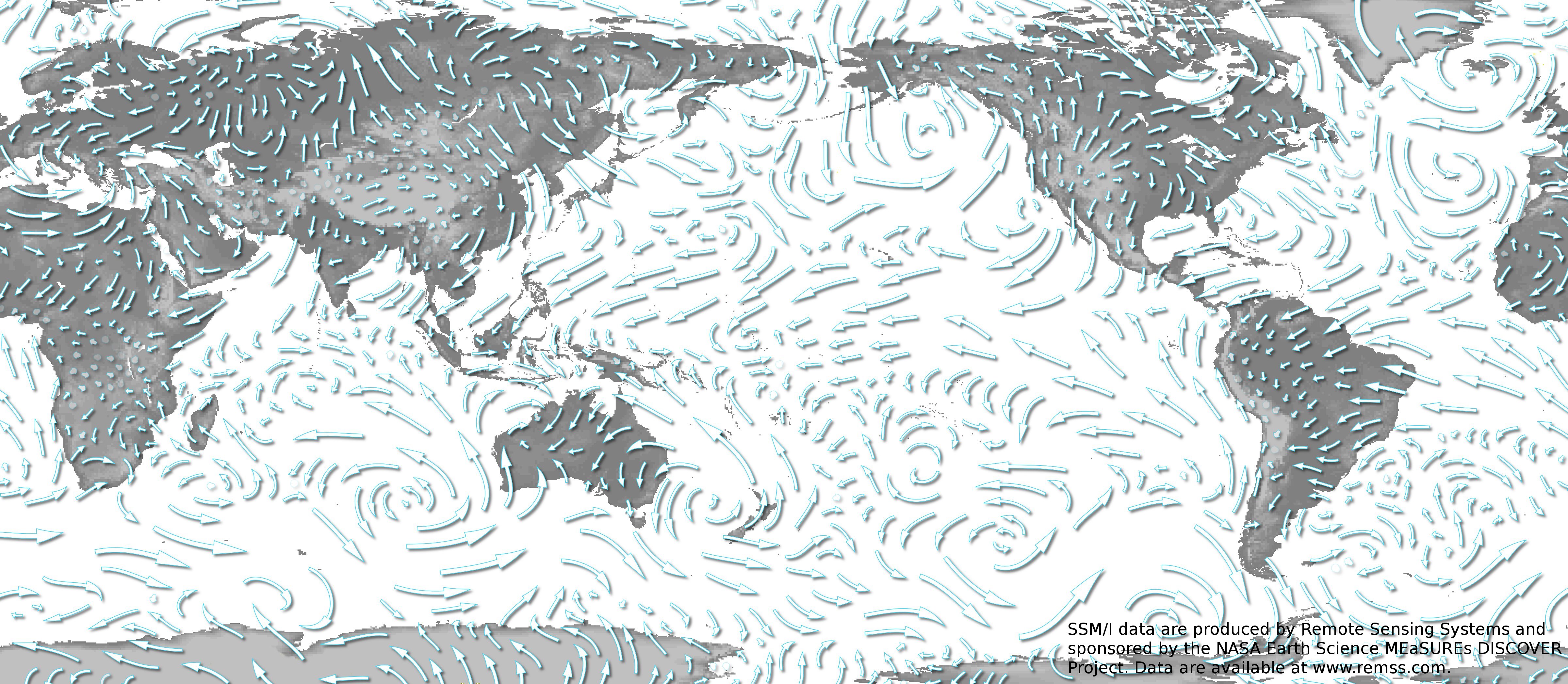}\\
\vspace{.020in}
\includegraphics[height=.355\linewidth]{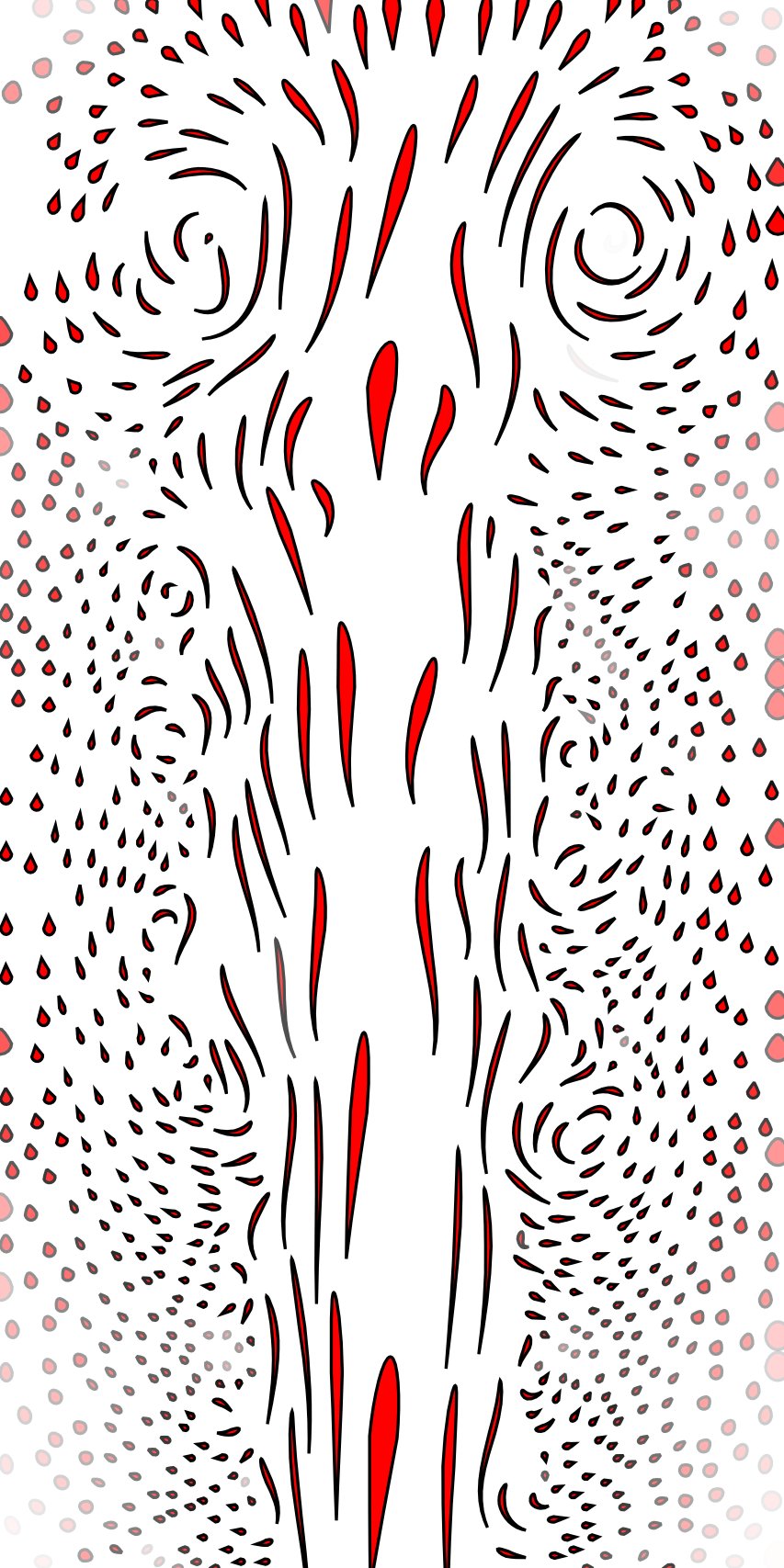}
\includegraphics[height=.355\linewidth]{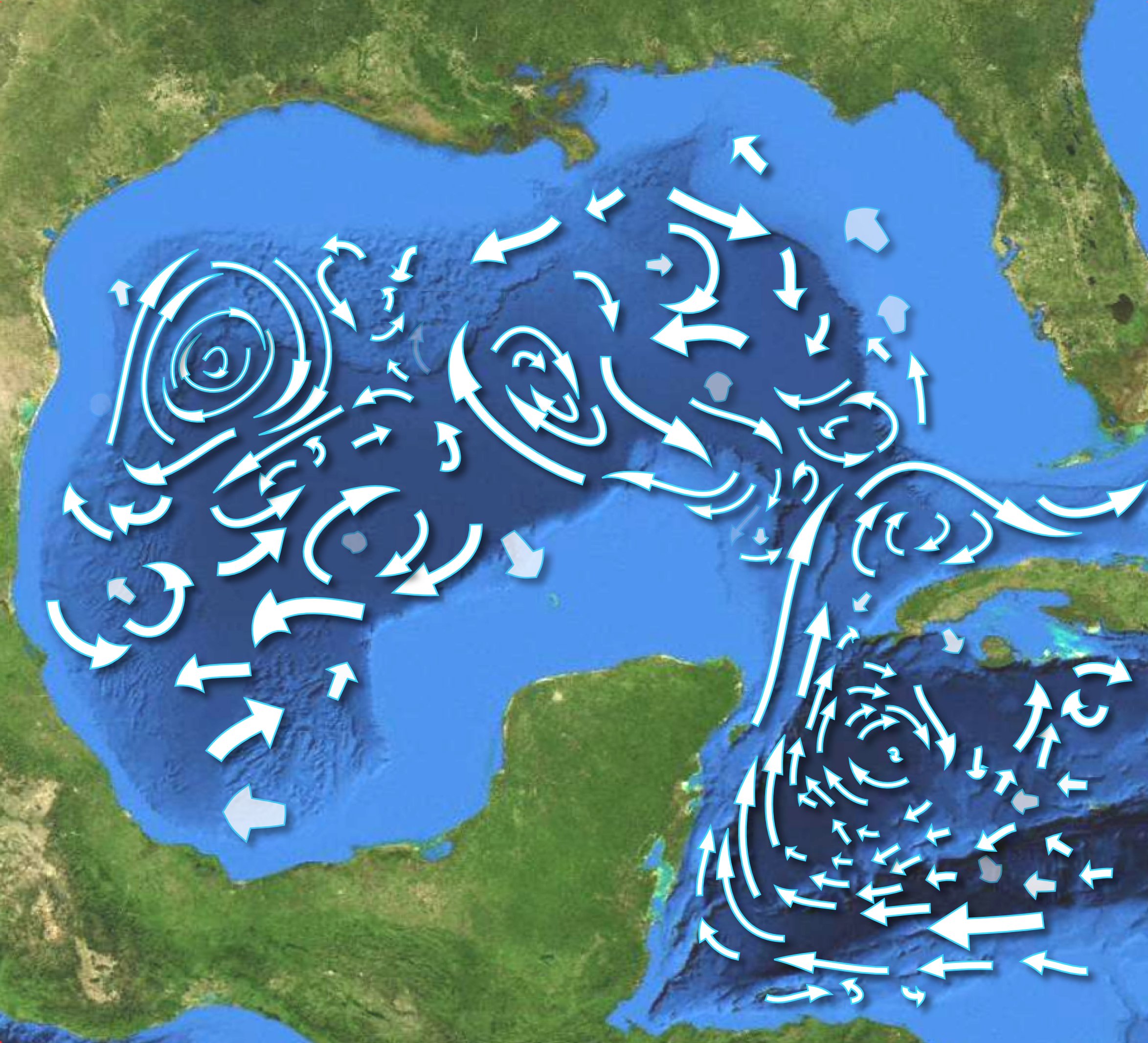}
\includegraphics[height=.355\linewidth]{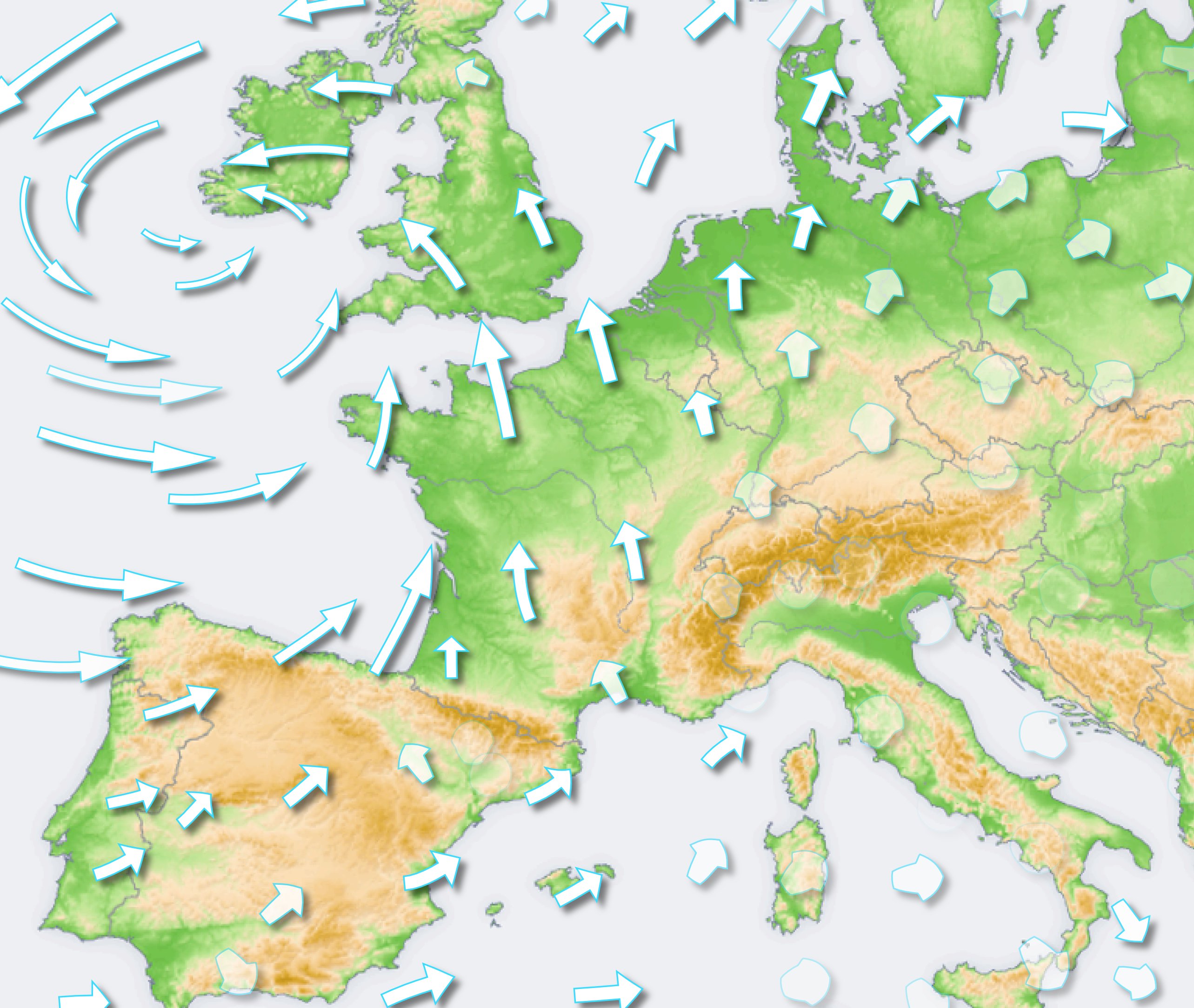}
\caption{\textbf{Frames of animated arrow plots of different datasets.}}
\label{fig:results}
\end{figure*}

We have tested our method on a water flow (gulf of Mexico), two winds data
(Europe's Storm in 1999 and Ocean winds) and a simulated velocity jet. Snapshots
of the resulting animations can be found in Figure~\ref{fig:results}, and videos
in the accompagning material. We tried to reflect the capabilities of our
method. The image of the 1999 Europe's storm shows coarse structure of the
cyclone, the arrows morph smoothly between zones with high and low velocity. The
flow in the Gulf of Mexico and the velocity jet have well established
currents as well as plenty of small turbulencies. To avoid overpopulation with
plenty of small arrows, we used high variation of underlying densities and thus
we capture tiny details while the well established flows are shown with longer
and thicker arrows.

\subsection*{Synthetic data}

\begin{figure}[!h]
\centering
\includegraphics[width=.45\linewidth]{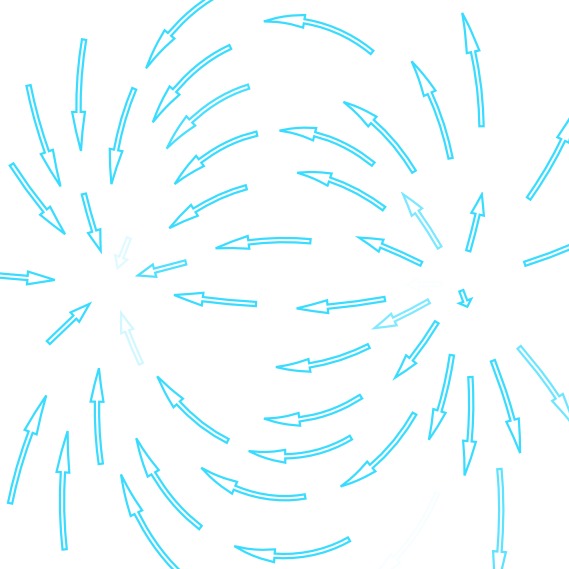}
\hspace{.1cm}
\includegraphics[width=.45\linewidth]{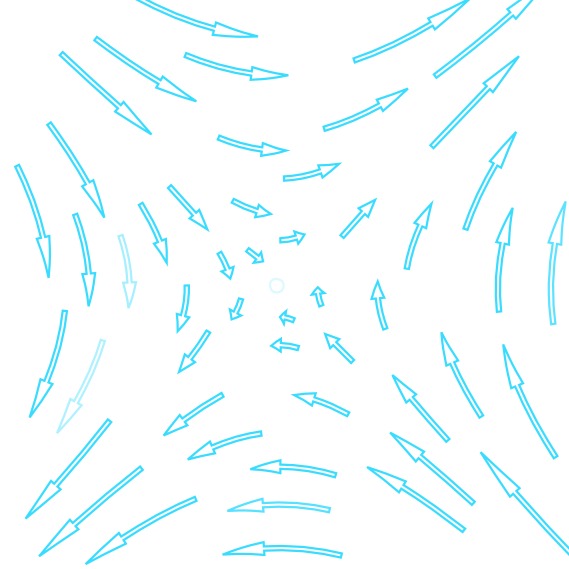}
\caption{\textbf{Stress tests} with synthetic vector fields having
extreme divergence.}
\label{fig:synthetic}
\end{figure}

The real data come from simulated or acquired flow fields, and they have a
limited divergence (due to the limited compressibility of the fluids). As a
consequence, the behavior of our method in extreme cases of divergence can not
be observed on such data, so we rely on synthetic fields to evaluate the
limitation of our method. Even in such cases, our algorithm is able to evenly
distribute arrows in the field, as illustrated in Figure~\ref{fig:synthetic}.

\subsection*{Timings}

The table below shows the time necessary to pre-compute various animations used
in this paper.

\begin{center}
\begin{tabular}{p{23mm}p{13mm}p{9mm}lc}
\hline
Dataset	& processing	& frames	& vector field \\
		& (seconds)		&			& resolution   \\
\hline
Storm dec 1999	& 4.1	& 48	& $385\times 325$	\\
Ocean~winds,~1987	& 56.7	& 40	& $1440\times 628$	\\
Velocity jet	& 45.5	& 500	& $128\times 256$	\\
Gulf of Mexico	& 19.25	& 183	& $352\times 320$ \\
Dipole	& 0.29	& 40	& $64\times 64$ \\
\hline
\end{tabular}
\end{center}

\section{Discussion}
\label{sec:discussion}

\subsection{Accuracy and resolution}

As stated in the introduction, a sparse set of arrows cannot represent all
details of the flow. To avoid high frequencies from perturbing the algorithm, it
is convenient to filter the data. A convolution with a Gaussian having the
minimal arrow size as standard deviation is sufficient to maintain undistorted
arrows. As a consequence, sampling the flow with a resolution higher than one
pixel corresponding to the minimal arrow size is not useful for the algorithm.

The time resolution of the flow do not impact the result quality as long as the
distance between neighbor arrows at each time step is a fair approximation of
this distance between two consecutive time steps. In practice, the displacement
of an arrow handle point between two frames should not be greater than the arrow
length.

\subsection{Streamlets vs. pathlets}
As mentioned in the section~\ref{sec:pbsettings}, the arrow shapes follow
streamlines, while their centers move on pathlines. This means that the arrows
in the visualization can point into different directions than the arrows move
to. It may seem more natural to use pathlets for arrow support. On the positive
side, arrows move into the direction they point to. On the negative side,
pathlets show history of the underlying field and not the current state of the
flow.

\begin{figure}[!htb]
\centering
\includegraphics[width=.48\linewidth]{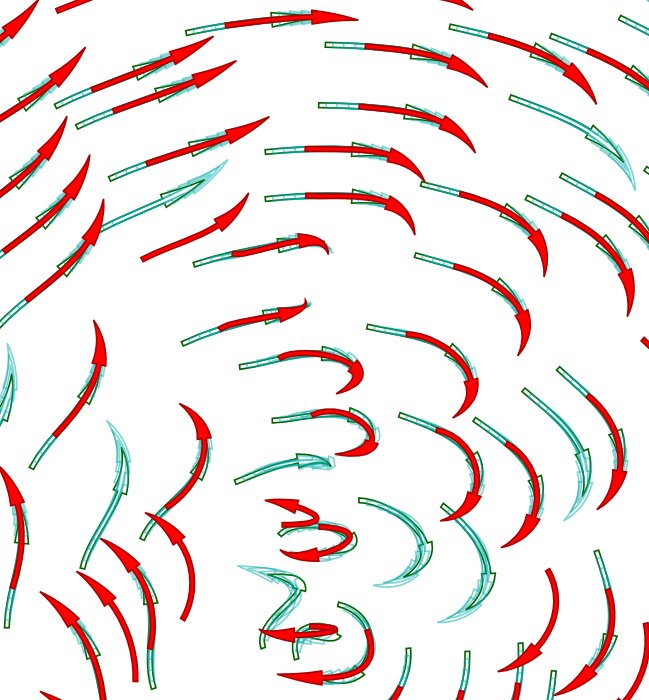}
\hspace{.1cm}
\includegraphics[width=.48\linewidth]{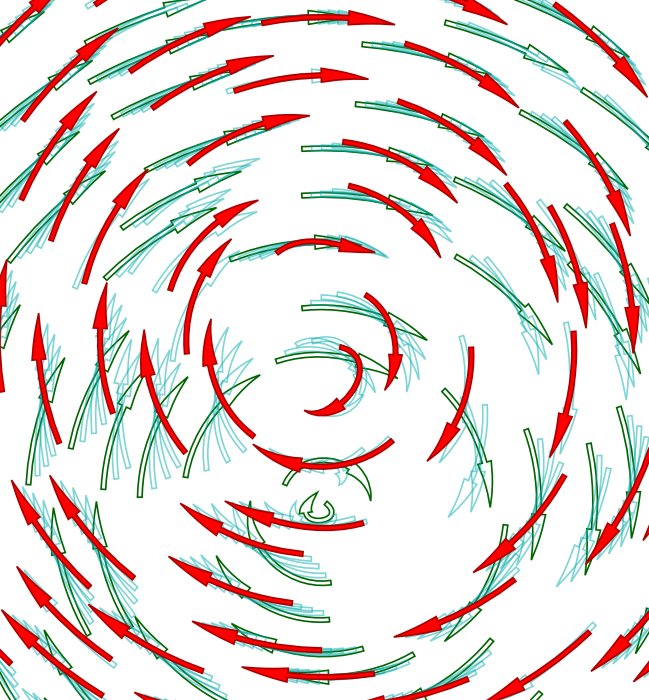}
\caption{\textbf{Arrows as pathlets or streamlets.} Both images represent the
same vortex moving from bottom to top. The arrows are bended along (Left)
pathlets and (Right) streamlets. Green hollow arrows show the field at timestep
$t-4$, red arrows at timestep $t$. Visually, the animation of streamlet-based
arrows is more intuitive than pathlet-based one.}
\label{fig:pathvsstream}
\end{figure}

In the case of steady flows or short arrows, using streamlet or pathlet is
almost equivalent. In other cases, using pathlines creates confusing effects as
illustrated in Figure~\ref{fig:pathvsstream}. Thus, we choose streamlets to
carry arrows in our animations.

\subsection{Comparison}
\label{sec:comparison_previous_works}

\begin{figure*}[!htb]
\centering
\includegraphics[width=.33\linewidth]{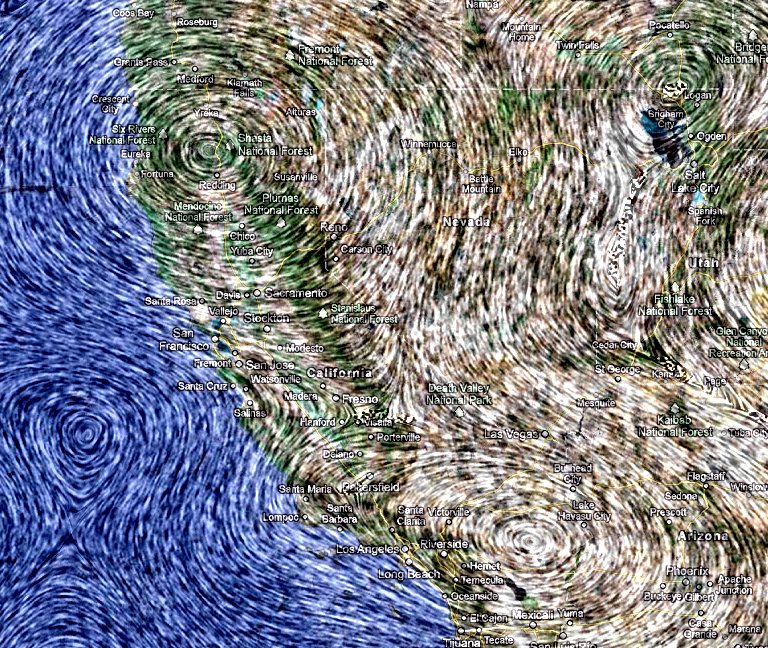}
\includegraphics[width=.33\linewidth]{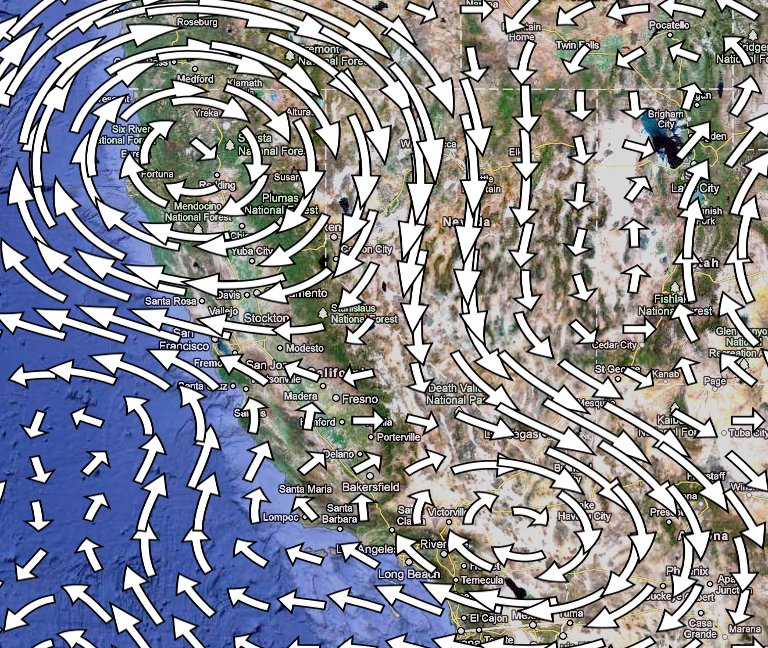}
\includegraphics[width=.33\linewidth]{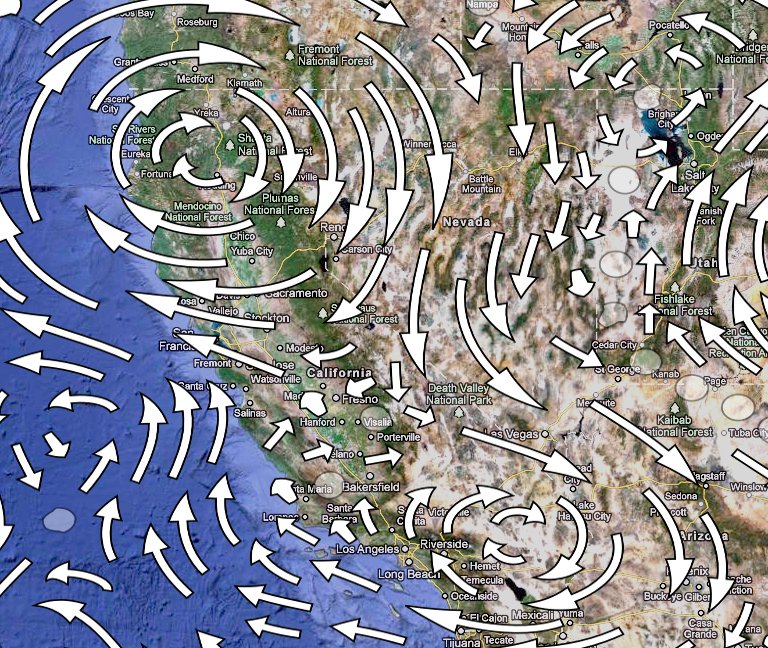}
\begin{minipage}[t]{.705\linewidth}
\vspace{0pt}
	\begin{minipage}[t]{.39\linewidth}
	\vspace{0pt}
	\includegraphics[width=\linewidth]{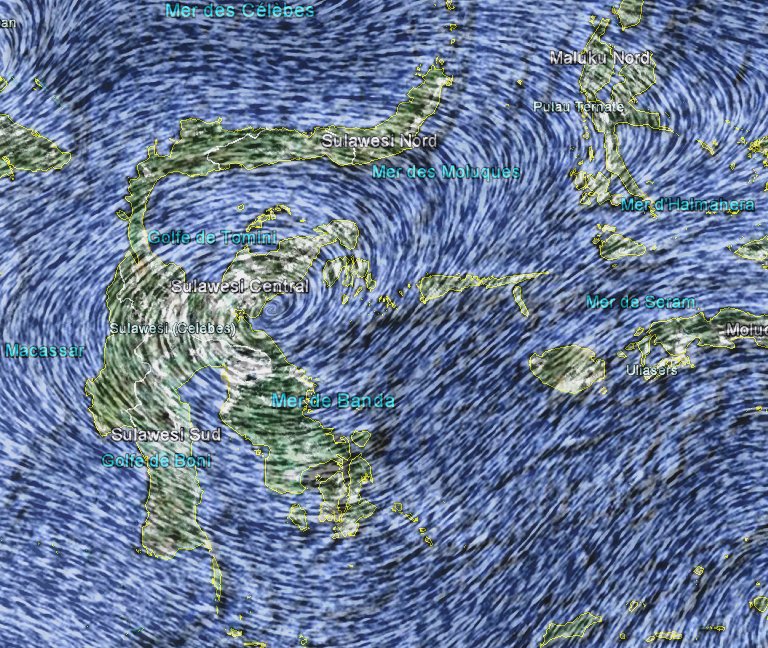}
	\end{minipage}
	\begin{minipage}[t]{.19\linewidth}
	\vspace{0pt}
	\includegraphics[width=\linewidth]{figures/indonesie0.jpg}\\
	\vspace{-0.32cm}
	\includegraphics[width=\linewidth]{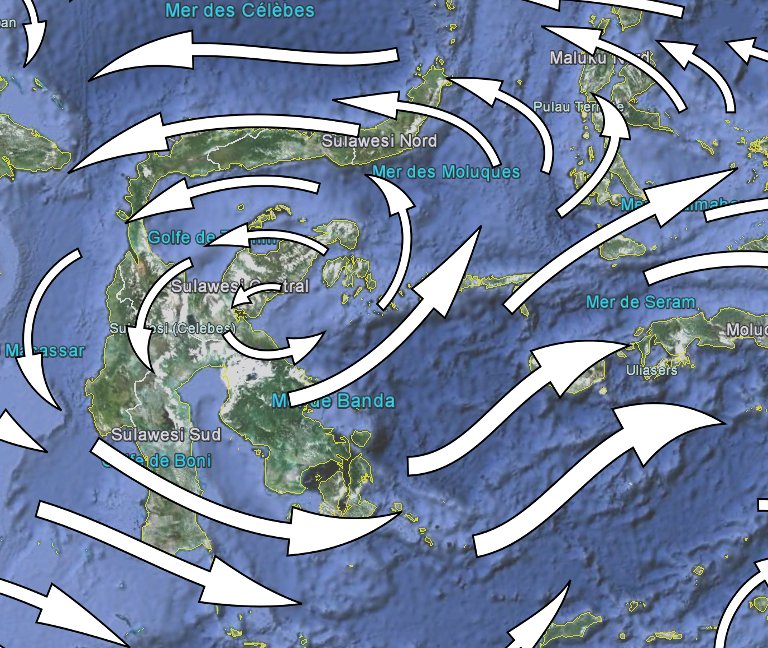}
	\end{minipage}
	\begin{minipage}[t]{.39\linewidth}
	\vspace{0pt}
	\includegraphics[width=\linewidth]{figures/indonesie1.jpg}
	\end{minipage}
\end{minipage}
\begin{minipage}[t]{.2875\linewidth}
\vspace{0pt}
\includegraphics[width=.48\linewidth]{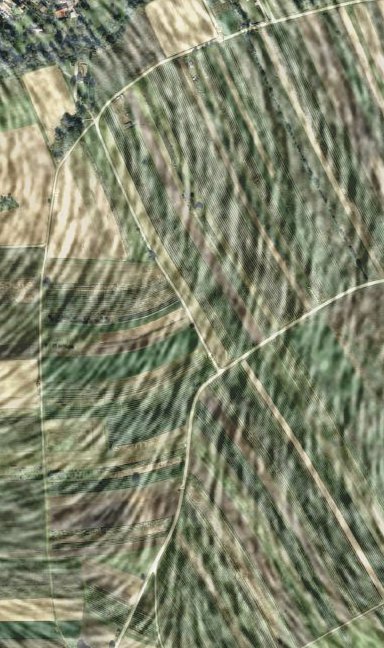}
\includegraphics[width=.48\linewidth]{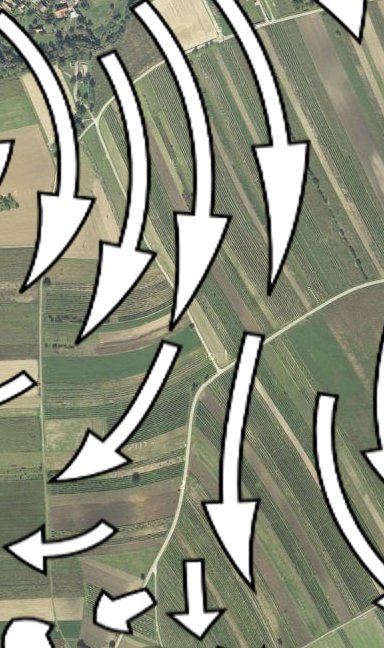}
\end{minipage}

\caption{\textbf{Visualizing a flow with a background image}. Top row,
left-to-right: Flow texture, bent arrows on a uniform grid and with our method.
Compared to flow textures, arrows temporary occlude the background but do
not deteriorate it. Bottom row, left images: arrow plots are less sensitive to the
display resolution; two right images: certain anisotropic textures such as
aerial photographies interfere with flow textures.}

\label{fig:comparison_lic}
\end{figure*}

As stated in the introduction, our objective is to produce pleasing and easy to
understand animations to represent a $2D$ flow field. As illustrated here with
flow textures and fixed position arrows, previous methods only partially satisfy
these constraints.

\textbf{Fixed position arrows} (see top middle image of
figure~\ref{fig:comparison_lic}) can overlap and their alignement can be
distracting. It is a fair solution to visualize fields in a realtime context, or
for vector fields where advection would be meaningless such as electromagnetic
fields that are not flows. However, our algorithm offers a more uniform coverage
and a natural displacement of the arrows.

\textbf{Flow texture} methods cannot represent flow orientation and
its magnitude without deteriorating global rendering quality. It is also
difficult to combine it with other sources of information such as a detailed
background as illustrated in figure~\ref{fig:comparison_lic}.

Moreover, texture based methods may suffer from bad rendering device (gamma or
resolution) and video compression. Our method does not provide an as accurate
representation of the flow details, but does not suffer from these drawbacks.

\section*{Acknowledgments}

We wish to thank MeteoSwiss for the dataset of the winds over Europe, the Center
for Ocean-Atmospheric Prediction Studies (COAPS) for the dataset of the ocean
currents in the Gulf of Mexico, Remote Sensing System for the dataset of the
winds around the world and Christoph Garth at UCDAVIS for the dataset of the
high velocity jet.

\section*{Conclusion}

We have developed a representation of dynamic vector fields by moving arrows.
Despite the intuition that the divergence will always create incessant popping
effects, we were able to produce convincing videos.
It was acheived by coupling arrow generation and rendering
strategies. Moreover, the extreme cases of divergence (source, sink) that
first come to mind as counter-examples that could challenge our method are
correctly handled by our algorithm and are not likely to occur often in real
flows.


\bibliographystyle{plain}
\bibliography{paper}

\end{document}